\documentclass{iopart}

\usepackage{latexsym}
\usepackage{amssymb}
\usepackage{cite}

\newcommand{\beq}{\begin{equation}}
\newcommand{\eeq}{\end{equation}}
\newcommand{\bea}{\begin{eqnarray}}
\newcommand{\eea}{\end{eqnarray}}
\newcommand{\barr}[1]{\begin{array}}
\newcommand{\earr}{\end{array}}
\newtheorem{theorem}{Theorem}[section]
\newtheorem{definition}{Definition}[section]

\newcommand{\bdf}{\begin{definition}}
\newcommand{\edf}{\end{definition}}
\newcommand{\bth}{\begin{theorem}}
\newcommand{\enth}{\end{theorem}}
\newcommand{\pd}{\partial}

\def\c+{\rlap{\ \raisebox{.2ex}{\scriptsize+}}\supset}
\begin{document}

\title{Are there Contact Transformations for Discrete Equations?}
\author{ Decio Levi~$^1$, Zora Thomova~$^2$ and Pavel Winternitz~$^3$}
\address{$^1$ Dipartimento di Ingegneria Elettronica, Universit\`a degli Studi Roma Tre 
and Sezione INFN, Roma Tre, Via della Vasca Navale, 84, 00146 Roma, Italy \\
$^2$ Department of Engineering, Sciences and Mathematics, SUNY Institute of Technology, 
100 Seymour Road, Utica, NY 13502, USA\\
$^3$ Centre de recherche math{\'e}matiques and D{\'e}partement de math{\'e}matiques et de statistique, 
Universit{\'e} de Montr{\'e}al, Case postale 6128, succ. centre-ville, Montr{\'e}al, Qu{\'e}bec, H3C 3J7, Canada}

\begin{abstract}

We define infinitesimal contact transformations for ordinary difference schemes as transformations 
that depend on $K+1$  lattice points $(K \geq 1)$ and can be integrated to form a local or global Lie group. 
We then prove that such contact transformations do not exist.
\end{abstract}

\maketitle

\section{Introduction}
Let us consider a system of ordinary or partial differential equations involving the independent variables $x \in \mathcal R^p$ and the dependent variables $y \in \mathcal R^q$.
Three types of transformation of variables play important roles in the theory of differential equations.

The first are point transformations, in which the new variables ($\tilde x, \tilde y$) depend only on the old variables ($x,y$). The methods of Lie point symmetry groups have been adapted to make them applicable to difference equations ( and differential -- difference equations \cite{doro1,lw1,doro2,lw2,lotw}.

The second are generalized transformations for which ($\tilde x, \tilde y$) depend not only on $x$ and $y$ but also on derivatives of $y$ with respect to $x$, like $y_{\alpha, x_i}$, $y_{\alpha, x_i x_k}$, $\cdots$ . These are important in the theory of integrable partial differential equations. They too have been adapted to difference equations and other equations involving discrete variables (or variables on lattices) \cite{lotw,Rav,MShRav}.

Intermediate between point and generalized transformations are contact ones. As symmetry transformations they provide non trivial results only for scalar equations \cite{ibraghimov84}
For them we have
\bea \label{1.1}
\tilde y = \Omega(x,y, y_x), \qquad \tilde x = \Lambda(x,y, y_x)
\eea
and a contact condition assures that
\bea \label{1.2}
\tilde y_x  = \Phi(x,y,y_x)
\eea
(no second derivatives).

The Lie algebra of contact transformations is closed under commutation, i.e. it does not introduce higher derivatives. This makes it possible to integrate the Lie algebra to a Lie group of transformations of the form (\ref{1.1}) and (\ref{1.2}). The inverse transformations have the same form. For some differential equations contact transformations provide analytical solutions, or at least a reduction of the equation, not obtainable by using point transformations \cite{hydon,im,yar}.

Contact transformations have, so far, not been adapted to difference equations. In this article we will just consider the case of ordinary difference equations. The purpose of this article is to analyze the situation in this case and show that contact transformations for ordinary difference equations do not exist.

In Section 2 we will review the theory of contact transformations in the case of ordinary differential equations while in Section 3 we review the theory of Lie point symmetries for difference schemes. Section 4 is then dedicated to proving that contact transformations for difference equations do not exist.
\section{Point and contact transformations for ordinary differential equations}

Let us consider an ODE of order $n$
\bea
y^{(n)} = F(x,y,y',\ldots, y^{(n-1)}) \label{e2-1}
\eea
and the n-th prologation of an element of its symmetry algebra
\bea
X= \xi \partial_x +\phi\partial_y, \quad pr^{(n)} X= \xi \pd_x + \phi \pd_y + \phi^{(1)} \pd_{y'} + \ldots + \phi^{(n)} \pd_{y^{(n)}}.  \label{e2-2}
\eea
The coefficients in the prolongation are calculated recursively
\bea
\phi^{(n)} = D_x \phi^{(n-1)} - y^{(n)} D_x \xi, \quad \phi^{(0)} = \phi \label{e2-4}
\eea
where $D_x$ is the total derivative operator.
The symmetry condition is
\bea
pr^{(n)} X (y^{(n)} - F) |_{y^{(n)}=F}=0. \label{e2-3}
\eea

For point transformations we have (by definition)
\bea
\xi=\xi(x,t) \qquad \phi = \phi(x,y) \label{e2-5}
\eea
and hence eq. (\ref{e2-4}) implies $\phi^{(n)} = \phi^{(n)} (x,y,y', \ldots , y^{(n)})$. A Lie algebra of point transformations can be integrated to a Lie group. Indeed we can integrate the vector field $X$
\bea
\frac{d \tilde{x}}{d \lambda} = \xi(\tilde{x}, \tilde{y}), \, \, \, \, \frac{d \tilde{y}}{d \lambda} = \phi(\tilde{x}, \tilde{y}) \label{e26} \\
\tilde{x}|_{\lambda =0} =x, \, \, \, \, \tilde{y}|_{\lambda =0} =y \nonumber
\eea

The transformation formulas for the derivatives $\tilde{y}_{k \tilde{x}} $ can be obtained (for any $k$) either by the chain rule, once $\tilde{y} (\tilde{x})$ is known, or by integrating $pr^{(k)} X$.

Now let us consider generalized symmetries containing higher order derivatives upto order $n$. In this case, instead of (\ref{e2-5}), we postulate
\bea
\xi =\xi(x,y,y', \ldots , y^{(n)}), \quad \phi=\phi(x,y,y', \ldots , y^{(n)}) \label{e2-7}
\eea
To be able to integrate the Lie algebra of such transformations to a Lie group, we need the coefficients
$\phi^{(k)}$ in the n-th prolongation (for $k=1,2,..., n$) to depend only on $x,y,y', ..., y^{(n)}$ but not on
$y^{(n+j)}$, $j \geq 1$.
H. Stephani \cite{STH} provides an elegant proof of the fact that this condition cannot be satisfied for $n >1$.

For $n=1$ contact transformations do exist if
\bea \label{e2-7a}
\phi_{y'} = y' \xi_{y'}
\eea
and an ODE can be invariant under a group of contact transformations.
The elements of Lie algebra have the form
\bea
 X= \xi(x,y,y') \pd_x + \phi(x,y,y') \pd_y + \phi^{(1)}(x,y,y') \pd_{y'}  \label{e2-8}
\eea
The contact condition (\ref{e2-7a}) eliminates $y''$ from $\phi^{(1)}$ and assures that $\phi^{(k)}$ for $k\geq 1$ is given by (\ref{e2-4}).  Thus first order contact transformations of the form (\ref{e2-8}) form a Lie algebra that can be integrated to give a Lie group of contact transformations.

\section{Ordinary difference schemes and their point transformation}

A difference equation on a fixed non-transforming lattice has very few continuous symmetries. One way of making a Lie symmetry approach to difference equations fruitful is to consider difference equations on flexible lattices that themselves transform under the transformations \cite{doro1,doro2,lw2}. Thus instead of difference equations we will deal with difference schemes.

Let us consider the case of one independent variable $x$ and one dependent one $y$. The independent variable $x$ is sampled at several points ${x_k}$ and $y(x)$ is evaluated at the same points $y_k = y_k(x_k)$. An ordinary difference scheme (O$\Delta$S) consists of two relations between the $K$ point ${x_k, y_k}$.
\bea
E_a ( \{ x_k \}, \{y_k \}, k=n+M, n+M+1, \ldots , n+N) =0 \label{e3-1} \\
a=1,2 \quad   K=N-M+1, \quad  n,M,N \in  \mathbf{Z}, \quad N>M \nonumber
\eea
In the continuous limit $x_{j+1}-x_j=h_{j+1} \rightarrow 0$ one of the two equations (\ref{e3-1}) should go into an ODE of order $K-1$ (or less), the other into an identity (like $0=0$). If the values ${x_k, y_k}$ are given in $K-1$ neighbouring points then we  must be able to calculate their values in the next point. The corresponding independence condition is
\bea
\frac{\pd (E_1,E_2)}{\pd (x_{n+N},y_{n+N})} \neq 0, \quad \forall n \label{e3-2}
\eea
Thus the O$\Delta$S (\ref{e3-1}) determines both the lattice and the solution of the difference equation.
The general solution of (\ref{e3-1}) can be written as
\bea
x_n &=& x(n,c_1,c_2,\ldots,c_{2(K-1)}) \label{e3-3} \\
y_n &=& y(n,c_1,c_2,\ldots,c_{2(K-1)}) = \phi_n(x_n,c_1,c_2,\ldots,c_{2(K-1)}) \nonumber
\eea

Point symmetries of the O$\Delta$S (\ref{e3-1}) can be determined algorithmically.
We require that the elements of the Lie point symmetry algebra of (\ref{e3-1}) have the form
\bea
X = \xi_n \pd_{x_n} + \phi_n \pd_{y_n} \label{e3-4}
\eea
with $\xi_n = \xi(x_n,y_n), \phi_n = \phi(x_n,y_n) $ and the prolongation of the vector field (\ref{e3-4})
should have the form
\bea
pr^{(K)} X = \sum_{k=n+M}^{n+N}  \left(  \xi_k \pd_{x_k} + \phi_k \pd_{y_k} \right) \label{e3-5}
\eea
where the sum is over all points figuring in the O$\Delta$S (\ref{e3-1}).
The determining equations are obtained from the conditions
\bea
pr^{(K)} X \, E_{a} |_{E_1 = 0, E_2=0} = 0   \qquad a=1,2  .  \label{e3-6}
\eea
Since each function $\xi_k$ and $\phi_k$ (for point transformations) only depends on one point $(x_k, y_k)$
equation (\ref{e3-6}) actually splits into a set of several functional equations for the functions
$\xi_k$ and $\phi_k$.

If we fix the value of $n$ then the $K$ points in (\ref{e3-1}) form a ``stencil'' \cite{doro2}.
Within a stencil we can choose different coordinates, more appropriate for taking a continuous limit.
Let us consider a stencil with the points $ \{ x_{n+k-1},y_{n+k-1}, 1 \leq k \leq K \} $ for some fixed $n$.
An alternative set of coordinates on the stencil is \cite{PW8}:
\bea
\{x_n,y_n,p_{n+1}^{(1)},p^{(2)}_{n+2},p^{(3)}_{n+3},p^{(4)}_{n+4}, \ldots p^{(K-1)}_{n+K-1},h_{n+1},h_{n+2},
\ldots h_{n+K-1}\}  \label{e3-7}
\eea
with
\bea
h_{n+k}&=&x_{n+k}-x_{n+k-1} \nonumber \\
p^{(1)}_{n+1}&=&\frac {y_{n+1}-y_{n}} {x_{n+1}-x_{n}} \label{e3-8} \\
p^{(2)}_{n+2}&= &2 \frac {p^{(1)}_{n+2}-p^{(1)}_{n+1}} { x_{n+2}-x_{n}} \nonumber \\
p^{(K-1)}_{n+K-1}&=& (K-1) \frac {p^{(K-2)}_{n+K-1}-p^{(K-2)}_{n+K-2}}{x_{n+K-1}-x_{n}}  \nonumber
\eea
In the continuous limit we have
\bea
h_{n+k} \rightarrow   0,  \, \,
p^{(1)}_{n+1} \rightarrow y', \, \,
p^{(2)}_{n+2} \rightarrow y'', \, \,
\ldots , \, \,
p^{(N)}_{n+K-1} \rightarrow y^{(K-1)}, \, \,  \label{e3-9}
\eea
If we transform the "discrete" prolongation (\ref{e3-5}) of the vector field (\ref{e3-4}) to the new
variables (\ref{e3-7}), we obtain
\bea
prX&=&\xi_n \pd_{x_n} + \phi_n \pd_{y_n} +
\sum_{k=1}^{K} \kappa^{(k)} \partial_{h_{n+k}} +
\sum_{k=1}^{K} \phi_{n+k}^{(k)} \partial_{p^{(k)}_{n+k}} \label{e3-10}
\eea
The general formulas for the coefficients $\kappa^{(k)}$ and $\phi^{(k)}$
are
\bea
\kappa^{(k)} &=& \xi_{n+k} - \xi_{n+k-1},      \label{e3-11}   \\
\phi^{(k)}_{n+k}&=&\Delta^{T} \phi^{(k-1)}_{n+k-1} - p^{(k)}_{n+k} \frac{1}{\sum_{j=1}^{k}  h_{n+j}}
\left( \sum_{j=1}^{k} h_{n+j}  \Delta^{T} \xi_{n+j-1}  \right) \quad
 k  =  1,2,\cdots , K  \nonumber
\eea
where the $\Delta^{T}$ is the total difference operator
\bea
\Delta^{T} F (x_n, y_n, p^{(1)}_{n+1},p^{(2)}_{n+2} ,\ldots)& = &
\frac{1}{h_{n+1}}
\left\{
F (x_{n+1}, y_{n+1}, p^{(1)}_{n+2},p^{(2)}_{n+3} \right. ,\ldots) \nonumber \\
& - &
\left. F (x_n, y_n, p^{(1)}_{n+1},p^{(2)}_{n+2} ,\ldots)
\right\} .
\eea

\section{Do contact transformations for O$\Delta$S exist?}

Let us again consider the O$\Delta$S (\ref{e3-1}).
The question that we pose is the following.
Is it possible to define contact transformations that leave the solution set
of the O$\Delta$S (\ref{e3-1}) invariant?
We shall impose the following restrictions to define contact transformations in the discrete case.

{\definition \label{d1} A contact transformation for the O$\Delta$S (\ref{e3-1}) satisfies:
\begin{enumerate}
\item
The vector fields forming the Lie algebra $L_c$ of contact and point transformations should have the form
(\ref{e3-4}) with at least one of the coefficients $\xi_n$ and $\phi_n$ in one of the fields $X$ depending on
$K+1$ points.
\bea
\xi_n &=& \xi_n (x_n, y_n, x_{n+1}, y_{n+1}, \ldots, x_{n+K}, y_{n+K} ), \, \, \, \, \, K \in \mathbb{Z}^{+} \nonumber \\
\phi_n &=& \phi_n (x_n, y_n, x_{n+1}, y_{n+1}, \ldots, x_{n+K}, y_{n+K} ) \label{e4-1}
\eea
\item
The Lie algebra should be integrable to a Lie group.
This implies that all coefficients (\ref{e3-11}) in the $K-th$ prolongation of $X$ (\ref{e3-10})
should depend only on the points $(x_n, y_n, \ldots, x_{n+K}, y_{n+K})$ and not on any further ones.
\item
In the continuous limit the algebra and the group of contact transformations of the O$\Delta$S (\ref{e3-1})
should reduce to the corresponding Lie algebra and Lie group of contact transformations of the corresponding ODE.
\end{enumerate}}

{\definition \label{d2} A contact transformation for a difference scheme is said to be of order $\ell$ if it involves $\ell+1$ points of the lattice.}

Let us now prove the following theorem:
{\theorem \label{t1}
Contact transformations of order 1 of O$\Delta$S do not exist.
}

{\bf Proof}

Let us consider the first prolongation of a vector field depending on two points:
\bea
pr^{(1)}X &=& \xi_n \pd_{x_n} + \phi_n \pd_{y_n} + \kappa^{(1)} \pd_{h_{n+1}} + \phi^{(1)}_{n+1} \pd_{p^{(1)}_{n+1}} \label{e4-2}
\eea
We have
\bea
\xi_n &=&  \xi_n(x_n, y_n, x_{n+1}, y_{n+1}) = \xi_n(x_n, y_n, x_{n}+h_{n+1}, y_{n}+h_{n+1}p^{(1)}_{n+1}) \nonumber \\
\phi_n &=&  \phi_n(x_n, y_n, x_{n+1}, y_{n+1}) = \phi_n(x_n, y_n, x_{n}+h_{n+1}, y_{n}+h_{n+1}p^{(1)}_{n+1}) \label{e4-3} \\
\kappa^{(1)} &=& \xi_{n+1} - \xi_n  \nonumber \\
\phi^{(1)}_{n+1} &=&
\frac{\phi_{n+1}-\phi_n}{h_{n+1}}
- p^{(1)}_{n+1}
\frac{\xi_{n+1}-\xi_n}{h_{n+1}},    \label{e4-4}
\eea
where
\bea
\xi_{n+1} &=&  \xi_{n+1}(x_{n+1}, y_{n+1}, x_{n+2}, y_{n+2}) \nonumber \\&=& \xi_{n+1}(x_{n}+h_{n+1}, y_{n}+h_{n+1}p^{(1)}_{n+1}, x_{n}+h_{n+1}+h_{n+2}, y_{n}+ \nonumber \\
&+&(h_{n+1}+h_{n+2})[p^{(1)}_{n+1}+\frac{h_{n+2}}{2} p^{(2)}_{n+2}]) \nonumber \\
\phi_{n+1} &=&  \phi_{n+1}(x_{n+1}, y_{n+1}, x_{n+2}, y_{n+2}) \label{e4-4a} \\&=& \phi_{n+1}(x_{n}+h_{n+1}, y_{n}+h_{n+1}p^{(1)}_{n+1}, x_{n}+h_{n+1}+h_{n+2}, y_{n}+ \nonumber \\
&+&(h_{n+1}+h_{n+2})[p^{(1)}_{n+1}+\frac{h_{n+2}}{2} p^{(2)}_{n+2}])  \nonumber
\eea
As one can see in (\ref{e4-4a})  both $\kappa^{(1)}$ and
$\phi^{(1)}_{n+1}$ depend  on $(h_{n+2},p^{(2)}_{n+2})$.
The question is whether this dependence can cancel out by proper choice of the symmetry generator form.
We have:
\bea
\frac{\pd \phi^{(1)}_{n+1}}{\pd p^{(2)}_{n+2}} =
\frac{h_{n+2} (h_{n+1}+h_{n+2})}{h_{n+1}} \left[
\frac{\pd \phi_{n+1}}{\pd y_{n+2}}- p^{(1)}_{n+1} \frac{\pd \xi_{n+1}}{\pd y_{n+2}} \right] = 0 ,  \label{e4-6}
\eea
which provide a contact condition similar to the continuous one (\ref{e2-7a}). However in this case we have to require the further condition
\bea
\frac{\pd \kappa^{(1)}}{\pd p^{(2)}_{n+2}} =
\frac{h_{n+2} (h_{n+1}+h_{n+2})}{2}
\frac{\pd \xi_{n+1}}{\pd y_{n+2}} = 0  \label{e4-5}
\eea
and hence
$\pd \xi_{n+1} / \pd y_{n+2} \, = \, 0$ and consequently from (\ref{e4-6}) $\pd \phi_{n+1} / \pd y_{n+2} \, = \, 0$.
In turn these conditions imply
\bea
\frac{\pd \xi_n}{\pd y_{n+1}} = \frac{\pd \phi_n}{\pd y_{n+1}} = 0 . \label{e4-7}
\eea
Similarly
$\pd \kappa^{(1)} / \pd h_{n+2} \, = \, 0$
and
$\pd \phi^{(1)}_{n+1} / \pd y_{n+2} \, = \, 0$
imply
\bea
\frac{\pd \xi_n}{\pd x_{n+1}} = \frac{\pd \phi_n}{\pd x_{n+1}} = 0 , \label{e4-8}
\eea
but (\ref{e4-7}) and (\ref{e4-8}) mean that (\ref{e4-2}) corresponds
to a point transformation, rather than a contact one.
\hfill $\Box$

By a similar calculation one can prove that if $\xi$ and $\phi$ depend on $K$ points then the $K-th$
prolongation will depend on one more point and hence one cannot obtain an integrable (closed)
system of contact transformations for an O$\Delta$S.
To sum up we state:

{\theorem \label{t2}
Contact transformation of any order $K \ge 1$ for O$\Delta$S do not exist.
}

If we consider discrete equations on a fixed (nontransforming) lattice then most, or all, point symmetries are absent. 
In this case the variable $x_n=hn+x_0$ with $h$ and $x_0$ fixed numbers and thus the infinitesimal generator is just of evolutionary form, i.e.
\bea \label{c-1}
X = \phi_n(y_n,y_{n+1})\partial_{y_n}+ \phi_{n+1}(y_{n+1},y_{n+2})\partial_{y_{n+1}} = \phi_n \partial_{y_n} + \phi_{n+1}^{(1)} \partial_{p^{(1)}_{n+1}} \label{eq5-37}
\eea
Then (\ref{eq5-37}) is a special case of (\ref{e3-10}) with $\xi_n$ and $\kappa^{(k)}$ absent. Thus the nontransforming lattice is a special case of the previous theorem.
\section{Conclusions}

Theorem \ref{t2} amounts to a ``no--go theorem''. It states that for ordinary difference 
schemes contact transformations, as defined in Definition \ref{d1}, do not exist. This leaves 
open the possibility that generalized symmetries might exist that in the continuous limit 
go into contact symmetries of an ODE. For comparison we recall that when considering integrable 
differential-difference equations on fixed (non-transforming) lattices the following situation occurs. 
An infinite Lie algebra of generalized symmetries exists of which a small subset  ``contracts"   
to point symmetries in the continuous limit \cite{hlrw}. Lie point symmetries of the corresponding 
differential equation are recovered in this manner.


\ack
We thank V.A. Dorodnitsyn, who participated in the early stages of this study, for helpful discussions.
DL  and ZT thank the CRM for the support during their visits to Montreal, where this research was carried out.
 DL  has been partly supported by the Italian Ministry of Education and Research, PRIN
``Nonlinear waves: integrable finite dimensional reductions and discretizations" from 2007
to 2009 and PRIN ``Continuous and discrete nonlinear integrable evolutions: from water
waves to symplectic maps" since 2010. The research of P.W. was partly supported by a research grant from NSERC of Canada.

\end{document}